\documentclass[11pt]{article}
\usepackage{moriond,epsfig}

\def\AJ{{\it Ap. J.} }
\def\AJL{{\it Ap. J. Lett.} }
\def\AJS{{\it Ap. J. Supp.} }

\def\ASAS{{\it Astron. and Astrophys.} }

\def\FP{{\it Fortschr. Physik} }
\def\GRG{{\it Gen. Relativity and Gravitation} }

\def\JP{{\it J. Phys.} }

\def\NAT{{\it Nature} }
\def\NC{{\it Il Nuovo Cimento} }
\def\NP{{\it Nucl. Phys.} }
\def\PL{{\it Phys. Lett.} }
\def\PR{{\it Phys. Rev.} }
\def\PRL{{\it Phys. Rev. Lett.} }

\def\PTP{{\it Progr. Theor. Phys.} }

\def\RMP{{\it Rev. Mod. Phys.} }

\def\al{\alpha}

\def\be{\begin{equation}}
\def\ee{\end{equation}}
\def\bea{\begin{eqnarray}}
\def\eea{\end{eqnarray}}

\def\al{\alpha}
\def\be{\beta}
\def\ga{\gamma}
\def\de{\delta}

\def\la{\lambda}

\def\Ga{\Gamma}
\def\De{\Delta}

 \def\frac#1#2{{\textstyle{{#1}\over
{#2}}}} 
\def\lsim{\mathrel{\rlap{\lower4pt\hbox{\hskip1pt$\sim$}}
    \raise1pt\hbox{$<$}}} \def\gsim{\mathrel{\rlap{\lower4pt\hbox{\hskip1pt$\sim$}}
    \raise1pt\hbox{$>$}}}
\def\sqr#1#2{{\vcenter{\vbox{\hrule height.#2pt
         \hbox{\vrule width.#2pt height#1pt \kern#1pt
         \vrule width.#2pt}
         \hrule height.#2pt}}}} \def\square{\mathchoice\sqr66\sqr66\sqr{2.1}3\sqr{1.5}3}

\def\beq{\begin{equation}}
\def\eeq{\end{equation}}
\def\beqa{\begin{eqnarray}} 
\def\eeqa{\end{eqnarray}}

\def\laq{\raise 0.4 ex \hbox{$<$}\kern -0.8 em\lower 0.62 ex\hbox{$\sim$}}
\def\gaq{\raise 0.4 ex \hbox{$>$}\kern -0.7 em\lower 0.62 ex\hbox{$\sim$}}

\begin{document}
\vspace*{3cm}
\title{THREE KEY TESTS TO GRAVITY}

\author{ORFEU BERTOLAMI}

\address{Instituto Superior T\'ecnico, Departamento de F\'\i sica,\\
Av. Rovisco Pais, 1; 1049-001 Lisbon, Portugal}

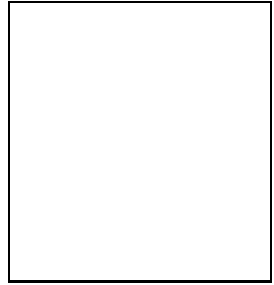
\maketitle\abstracts{
Three presumably 
unrelated open questions concerning gravity and the structure of the 
Universe are here discussed: 1) To which extent is Lorentz invariance 
an exact symmetry ? 2) What is the equation of state of the Universe ?  
3) What is the origin of the so-called Pioneer anomaly ?  
}

\section{Lorentz Symmetry}

Lorentz invariance is one of the most well established 
symmetries of physics and is a basic ingredient in  
all known physical theories. However, recently, there has been evidence 
that this symmetry may be broken in at least three different phenomena:

\vspace{0.2cm}

\noindent
i) Observation of ultra-high energy cosmic rays (UHECRs) with energies 
\cite{Hayashida} 
beyond the so-called Greisen-Zatsepin-Kuzmin (GZK) 
cutoff \cite{Greisen}, $E_{GZK} \simeq 4 \times 10^{19}~eV$. Breaking 
of Lorentz invariance may explain these events as it implies that
resonant scattering reactions with photons of the Cosmic Microwave 
Background Radiation (CMBR), e.g. 
$p + \ga_{2.73K} \to \De_{1232}$, are suppressed 
\cite{Sato1,Coleman,Bertolami1}. Astrophysical solutions 
to this paradox are 
possible and require identifying energetically viable sources within \cite{Stecker,Hill}
$D_{Source} \lsim 50 - 100~Mpc$, so that the traveling time of the 
emitted particles 
is shorter than the attenuation time due to particle photoproduction on the 
CMBR. Within a volume of radius 
$50 - 100~Mpc$ around us only neutron stars, active galactic 
nuclei, gamma-ray bursts and cluster of galaxies 
are feasible acceleration sites \cite{Hillas,Cronin}. However, problems related with 
the lack of spatial correlations 
between observed UHECRs and candidate sources, and the mismatch of fluxes has
rendered these alternatives untenable so far
(see e.g. Ref. [10] and references therein).

\vspace{0.2cm}

\noindent
ii) Events involving gamma radiation with energies beyond
$20~TeV$ from distant sources such as Markarian 421 ($z = 0.031$) and 
Markarian 501  ($z = 0.033$, $D \simeq 157~Mpc$) blazars
\cite{Krennrich}. These observations can be explained via the 
breaking of Lorentz symmetry as otherwise, 
due to pair creation, there should exist a strong attenuation of fluxes beyond $100~Mpc$ 
by the diffuse extragalactic background of infrared photons \cite{Amelino1}.

\vspace{0.2cm}

\noindent
iii) Studies of the
evolution of air showers produced by ultra high-energy hadronic particles 
suggest that pions live longer than expected \cite{Antonov}. 

\vskip 0.3cm

\noindent
Of course, breaking of Lorentz symmetry may lead to other threshold effects 
related with pair creation with asymmetric momenta, photon stability, alternative  
\v{C}erenkov effects, etc \cite{Konopka,Jacobsen}.
 
From the theoretical point of view, work in the context of string/M-theory shows 
that Lorentz symmetry may be spontaneously broken by
nontrivial solutions in string field theory  \cite{Kostelecky1}, a feature that is also found
in loop quantum gravity \cite{Gambini,Alfaro}, in noncommutative field theories 
\cite{Carroll,Bertolami5}, and in quantum gravity inspired spacetime foam 
scenarios \cite{Garay}. A related issue arising in noncommutative field theories 
concerns the breaking of translational symmetry by interactions, leading to a potential solution 
for the abovementioned astrophysical problems \cite{Bertolami15}.   
The interactions resulting from the breaking of Lorentz symmetry may lead to 
striking implications at low-energy 
\cite{Bertolami6,Bertolami3,Mavromatos,Aloisio} and as well as to
the breaking of CPT symmetry \cite{Kostelecky2}. The breaking of CPT allows, on its hand, 
for a thermodynamical mechanism for baryogenesis \cite{Bertolami4}. 
An extension of the Standard Model (SM) that incorporates violations 
of Lorentz and CPT 
symmetries has already been built \cite{Colladay2}.

On quite general terms, attempts to explain the three discussed paradoxes rely on  
deformations of the relativistic dispersion relation for a particle species  
$a$, such as

\begin{equation}
E_a^2 = p_a^2 c_a^2 + m_a^2 c_a^4 + F(E_a, p_a, m_a, c_a)
\quad,
\label{1}
\end{equation}    
where $c_a$ is the maximal attainable velocity for particle $a$ and $F$ 
is a function of $c_a$ and of the relevant 
kinematic variables. 

For example, Coleman and Glashow \cite{Coleman} consider, to explain the observation 
of cosmic rays beyond 
the GZK cutoff, that each particle has its own maximal attainable velocity and $F = 0$. 
A tiny difference between the maximal 
attainable velocities, $c_p - c_{\Delta} \equiv \epsilon_{p \Delta} 
\simeq 1.7 \times 10^{-25}~c$, can explain the events beyond the GZK 
cutoff \cite{Coleman}. This bound on Lorentz symmetry is about three orders of magnitude 
more stringent than the experimental one \cite{Lamoreaux}, 
$\delta < 3 \times 10^{-22}$.

Function $F$ arising from the Lorentz violating extension of the SM fermionic sector discussed in 
Ref. [29] is shown to be \cite{Bertolami1}:

\begin{equation}
F = -2 c_{00}E^2 \pm 2 d_{00}Ep
\quad, 
\label{4}
\end{equation}
with $c_a = c$ for all particles. Coefficients $c_{00}$ and $d_{00}$ are time-like and 
flavour-dependent, while the plus/minus sign concerns the difference in the chirality of particles.

It has been pointed out that some quantum gravity,
stringy inspired and noncommutative field theory models lead to a modification of 
the dispersion relation that is cubic in the momentum \cite{Alfaro,Bertolami15,Mavromatos,Aloisio}:

\begin{equation}
F = - k_a {p_a^3 \over M}
\quad, 
\label{5}
\end{equation}
where $k_a$ is a constant and $M$ a mass scale. A deformation of this type can potentially 
explain the three discussed paradoxes \cite{Konopka,Alfaro,Mavromatos,Aloisio,Amelino2}. 
On any account, it is clear that further 
observational and theoretical work is still required in order to settle to which extent 
Lorentz symmetry is an exact symmetry. Implications for gravity can be manifold. We shall 
discuss in Section 3 an example.

\section{Generalized Chaplygin Gas Model}

It has recently been suggested that evidence for a dark energy
component to the total energy density of the Universe as inferred from Type Ia 
Supernovae observations \cite{Perlmutter}
might be accounted by a change in the equation 
of state of the background fluid \cite{Kamenshchik}, rather than by a cosmological constant
or the dynamics of a scalar field rolling down a potential. 
In the context of Friedmann-Robertson-Walker cosmology, one considers   
an exotic background fluid, the generalized Chaplygin gas (GCG), 
which is described by the equation of state

\beq
p_{ch} = - {A \over \rho_{ch}^\alpha}~~,
\label{eq:eqstate}
\eeq
\vskip 0.3cm

\noindent
where $\alpha$ is a constant in the range 
$0 < \alpha \le 1$ (the Chaplygin gas corresponds to the case $\alpha=1$, 
and clearly $\alpha = 0$ to the $\Lambda$CDM model) and
$A$ a positive constant. Inserting this equation of state 
into the relativistic energy conservation equation, leads to an energy 
density that evolves as \cite{Bento1}

\beq
\rho_{ch} =  \left(A + {B \over a^{3 (1 + \alpha)}}\right)^{1 \over 1 +
 \alpha}~~,
\label{eq:rhoc}
\eeq 
\vskip 0.3cm

\noindent
where $a$ is the scale-factor of the Universe and $B$ an integration 
constant. It is remarkable that the model naturally interpolates between 
a universe dominated by dust at early time and a De Sitter one at late time, with 
an intermediate a phase described by a cosmological constant plus
matter with a ``soft'' equation of state, $p = \alpha \rho$ ($\alpha \not= 1$).
Notice that the chosen interval for $\alpha$ ensures that 
the sound velocity ($c_s^2 = \alpha A/ \rho_{ch}^{1+\alpha}$) 
does not exceed,
in the ``soft'' equation of state phase, 
the velocity of light. Furthermore, as pointed out in Ref. [34], 
it is only for $0 < \alpha \le 1$ 
that the analysis of the evolution of energy density fluctuations makes sense. 
It is also shown that the GCG can be described
by a complex scalar field with a 
generalized Born-Infeld action \cite{Bento1}. 
It is clear that the GCG is a potential candidate for explaining the observed
accelerated expansion of the Universe as it  
leads to an asymptotic phase where the equation of state is dominated by a 
cosmological constant, $8 \pi G A^{1/1+\alpha}$. 
It has also been shown that the model admits, under conditions, 
an inhomogeneous generalization which can be regarded as a {\it unification 
of dark matter and dark energy} \cite{Bento1,Bilic} without conflict with 
standard structure formation
scenarios \cite{Bento1,Bilic,Fabris}. Thus the GCG model is 
an interesting alternative to models where the 
accelerated expansion of the Universe is 
due to an uncanceled cosmological constant (see [42] and 
references therein) or a rolling scalar field as in quintessence
models with one \cite{quint1} or two scalar fields \cite{quint2}.

The possibility of describing dark energy via the GCG 
has led to great interest in 
constraining the model using  observational data, 
particularly those arising from Type Ia Supernovae 
\cite{Supern,Makler} and gravitational lensing statistics 
\cite{Silva}. Furthermore, it has been shown
more recently that 
the positions of peaks and troughs of the CMBR power spectrum as arising from WMAP and BOOMERanG data, 
allow constraining a 
sizeable portion of the parameter space of the GCG model \cite{Bento4,Bento5} (see also Refs. [45,46]).
Indeed, bounds on the locations of the first two  peaks and the first
trough,
from  WMAP measurements of the CMBR temperature angular power spectrum
\cite{WMAP}:

\beqa
\ell_{p_1} &=& 220.1\pm 0.8~~,\nonumber\\  
\ell_{p_2} &=& 546\pm 10~~,\nonumber\\  
\ell_{d_1} &=& 411.7\pm 3.5~~,   
 \label{eq:wmap}
\eeqa
and the location of the third peak, from BOOMERanG measurements 
\cite{Boomerang}

\beq
\ell_{p_3} = 825^{+10}_{-13}~~,
\label{eq:l3}
\eeq
do imply in nontrivial constraints on the parameters of the GCC model.

These constraints can be summarized as follows \cite{Bento5}. 
The Chaplygin gas model, 
$\alpha = 1$, is incompatible
with the data and so are models with $\alpha \gsim 0.6$. For $\alpha = 0.6$, 
consistency 
with data requires for the spectral tilt, $n_s > 0.97$. 
Actually, $\Lambda$CDM model barely fits the 
data for $n_s \simeq 1$ (notice that  WMAP data yield 
$n_s=0.99\pm0.04$) and $h\gsim 0.71$ is required ($H_0 = 100 h~kms^{-1}Mpc^{-1}$); 
for lower values of $n_s$, $\Lambda$CDM  and the GCG model give similar results. 
Moreover, one finds that data favours $\alpha \simeq 0.3$ GCG models.

These results are compatible with the ones of Ref. [43] using BOMERanG data for the 
third peak and Archeops \cite{Benoit} data for the first peak, as well as Type Ia 
Supernovae bounds,
namely $0.2 \lsim \alpha \lsim 0.6,~0.81\lsim A_s \lsim 0.85$, where 
$A_s \equiv A/ \rho_{ch0}^{1+\alpha}$. 

As pointed out in Ref. [43], considering, besides CMBR and Supernovae data, 
bounds arising from faraway quasar sources lead to more stringent 
bounds on the parameters of the GCG model. 
Bounds arising from Type Ia Supernovae data, which suggest that \cite{Makler}
$0.6 \lsim A_s \lsim 0.85$, are consistent with results 
of Ref. [44] for
 $n_s=1$ and $h=0.71$. 

It worth mentioning that future Supernovae data together with 
statistics of gravitational lensing surveys may further
constrain the parameter space of the GCG model,
although such studies are very dependent on the considered fiducial
models \cite{Silva}. 

The interested reader is referred to Refs. [43,44] 
for technical details and to Ref. [50] for a brief review on 
the GCG model.

\section{Pioneer Anomaly}

Studies of radiometric data from the Pioneer 10/11, Galileo and
Ulysses have revealed the existence of an anomalous acceleration on
all four spacecraft, inbound to the Sun and with a constant
magnitude of $a_A \simeq (8.5 \pm 1.3) \times 10^{-10}~ m s^{-2}$.
Attempts to explain this anomaly as a result of poor
accounting of thermal and mechanical effects, as well as errors in
the tracking algorithms, have shown to be unsuccessful
\cite{Nieto1} (see however Ref. [52]).
Since the two Pioneer spacecraft follow approximate opposite hyperbolic
trajectories away from the Solar System, while Galileo and Ulysses
describe closed orbits, and from the fact that the
three designs are geometrically distinct, it seems plausible to assume that 
this anomaly involves new physics.

Many proposals have been advanced to explain this anomaly: a 
Yukawa-like or
higher order corrections to the Newtonian potential
\cite{Anderson}, a scalar-tensor extension to the standard
gravitational model \cite{Calchi}, interaction of the spacecraft
with a long-range scalar field, determined
by an external source term proportional to the
Newtonian potential \cite{Mbelek}, to mention just a few.
It is interesting that in higher-curvature theories of gravity 
where the gravitational coupling 
is asymptotically free, a feature that has attracted some attention 
in the context of the dark matter problem 
\cite{Goldman,Bertolami10,Bertolami11}, the 
gravitational coupling is stronger on large scales, thus implying, 
in principle, to a Pioneer-like anomalous acceleration.

As an alternative explanation, we exploit the implications of the 
idea that the Pioneer anomaly
reflects the bimetric nature of spacetime in the Solar System 
\cite{Bertolami12}.
This proposal, although not fully consistent (see below), 
is based on the bimetric theory of gravity put forward 
long ago by Rosen \cite{Rosen}.

One considers a region of spacetime endowed, 
besides the dynamical metric $g_{\mu\nu}$, with a non-dynamic
metric $\eta_{\mu\nu}$, that has signature $+2$, is Riemann flat
or has a constant curvature. The
background metric $\eta_{\mu\nu}$ is static and
asymptotically $g_{\mu\nu} + n_{\mu\nu} \rightarrow n_{\mu\nu}$.

In this theory the equation for the dynamical gravitational field is given by 
\cite{Rosen}:

\beq 
\square_\eta g_{\mu\nu} - g^{\al\be}\eta^{\ga\de}g_{\mu\al |
\ga}g_{\nu\be |\de} = -16 \pi G (g/\eta)^{1/2}(T_{\mu\nu}-{1 \over
2} g_{\mu\nu} T)~~, 
\label{fieldeqs} 
\eeq

\noindent 
where $T \equiv T_{\mu\nu} g^{\mu\nu}$ is the trace of the 
energy-momentum of matter and
$\square_\eta$ is the D'Alembertian operator with respect to
$\eta_{\mu\nu}$. 
One can always choose coordinates in which $\left( \eta_{\mu\nu}
\right) = diag(-1,1,1,1)$, and $ \left(
g_{\mu\nu} \right) = diag(-c_0,c_1,c_1,c_1)$, where $c_0$ and
$c_1$ are parameters that may vary on a Hubble $H^{-1}$ timescale
\cite{Rosen}.

Rosen's bimetric theory explicitly breaks Lorentz invariance. This can be
understood by resorting to the Parametrized Post-Newtonian (PPN)
formalism, a systematic expansion of first-order $1/c^2$ terms in
the Newtonian gravitational potential and related quantities
\cite{Will1}. It turns out that all metric theories of gravity
can be classified according to ten PPN parameters:
$\ga,\be,\xi,\al_1,\al_2,\al_3,\zeta_1,\zeta_2,\zeta_3,\zeta_4$.
These are the linear coefficients of each possible first-order
term (generated from rest mass, energy, pressure and velocity),
and relate a particular theory with fundamental aspects of
physics: conservation of linear and angular momentum,
preferred-frame and preferred-location effects, nonlinearity and
space-curvature per unit mass, etc.

Einstein's General Relativity, the most successful theory up
to date, exhibits a set of PPN parameters where $ \be = \ga = 1$ and
the remaining ones are equal to zero. Rosen's bimetric theory has 
$\be = \ga = 1 $, but a nonvanishing $\al_2 = c_0/c_1 - 1$
coefficient. This indicates that the theory is semiconservative, i.e. 
angular momentum is not conserved, and that it exhibits preferred-frame effects, 
meaning that Lorentz symmetry is broken and the Strong Equivalence Principle
does not hold.    

By linearizing Eq. (\ref{fieldeqs}) in the vacuum, one obtains the
wave equation for weak gravitational waves, whose solution corresponds to a
wave propagating with speed $c_g = \sqrt{c_1/c_0}$. Thus, $\al_2$
measures the difference in the propagation velocities (measured by an observer
at rest in the Universe rest frame) between electromagnetic and
gravitational waves \cite{Will1}.
A rigorous study of deviations between the Sun's spin axis and the
ecliptic does lead to the experimental constraint \cite{Nordtvedt} $|\al_2| < 1.2
\times 10^{-7}$. 

Due to their small mass, the abovementioned spacecraft can be regarded test
particles, and their acceleration can be obtained by
computing the timelike geodesics of the full metric,
$h_{\mu\nu} = \eta_{\mu\nu} + g_{\mu\nu} $:

\beq 
{d^2 x^\mu \over d \tau^2}+ \Ga^\mu_{\al\be}{d x^\al \over d
\tau}{d x^\be \over d \tau}=0~~. 
\label{tgeodesic} 
\eeq

\noindent 
In the Newtonian limit and for $v \ll c$, one obtains for a 
diagonal metric $h_{\mu \nu}$ the acceleration

\beq 
{a^i} \simeq -\Ga_{00}^i = {1 \over 2} h^{i \la}
\partial_\la h_{00} ~~. 
\label{accel} 
\eeq

In the case of a very weak central gravitational field, such as in
the Solar System, the background metric is not flat, but given (in
Cartesian coordinates with the Sun in the origin) by 
$(g_{\mu \nu}) = diag(-1 - 2U, 1,1,1)$,
where $U(r) = - G M_\odot / r \equiv - C / r $ is the
gravitational potential. Substitution into Eq. (\ref{accel}) yields

\beq 
a_i = -( 1 + c_1 )^{-1} \left[ \partial_i U + {1 \over 2}
\partial_i c_0 \right] ~~,
\eeq
from which one can identify the radial anomalous component of the
acceleration:

\beq 
\vec{a}_A = c_1 \vec{\nabla}_r U - {(1 - c_1) \over 2}~
\vec{\nabla}_r c_0 ~~.
\label{anomalous} 
\eeq

It can be readily seen that, if $c_0$ and $c_1$ are
homogeneous in space, the derived anomalous acceleration is not
constant, which seems to contradict the observation. Therefore, 
assuming that these parameters depend on the distance to the Sun,
the simplest choice consistent 
with an homogeneous $\al_2$ is:

\beq 
c_0=A~r~~,~~c_1=B~r 
\label{ansatz} ~~,
\eeq 
with $ A, B > 0 $ so that the resulting anomalous
acceleration is inbound. Clearly, from the bound on $\al_2$, $|A/B-1| <
4 \times 10^{-7} $ it follows that $ A \simeq B $. Notice that $A$ cannot
be exactly equal to $B$, since this implies that $\al_2=0$,
meaning that General Relativity is recovered, as there is a coordinate transformation
which pulls the metric $g_{\mu\nu}$ back to the asymptotically
flat metric $\eta_{\mu\nu} $.

Substituting {\it Ansatz} Eq. (\ref{ansatz}) into 
Eq. (\ref{anomalous}), leads to

\beq 
a_A = -{A \over 2} \left[ 1 - {B \over A}{2C \over r} - B r
\right] \simeq -{A \over 2} \left[1 - {2C \over r} - A r \right]
\label{anom}~~. 
\eeq

\noindent 
In geometric units ($c=G=1$), $|a_A| \simeq 8.5 \times 10^{-10}~ m s^{-2} 
= 1.5 \times 10^{-15}~AU^{-1}$, from which follows that $ A = 3 \times
10^{-15}~AU^{-1}$ as $C \equiv G M_\odot = 10^{-8}~ AU$.

Hence, in this proposal 
an hypothetical dedicated probe to confirm the Pioneer anomaly, as 
suggested in Refs. [63,64], would not need to venture into too deep space
to detect such an anomalous acceleration, but just to a distance
where it is measurable against the regular acceleration and the
solar radiative pressure, actually approximately from Jupiter onwards. 
Unfortunately, as it stands, Rosen's bimetric theory, is inconsistent, 
as it predicts emission of dipolar gravitational radiation from 
gravitationally bound systems, which is incompatible with data from the binary \cite{Will2} 
PSR 1913+16, or no gravitational radiation at all \cite{Rosen1}. 
Possible alternatives will be presented elsewhere \cite{Bertolami12}. In any case, 
it seems fair to conclude that if the Pioneer anomaly turns out to be a real physical 
effect it may be an important clue on new gravitational physics. Unfortunately, given 
that Pioneer 10 has, since last February, fallen beyond detectability, most likely 
only through 
a dedicated mission the desired confirmation will be achieved \cite{Bertolami,Anderson2}.

\section*{Acknowledgments}

This contribution is based on work developed with various 
collaborators and friends. I would like to mention their names and thank them explicitly: 
Maria Bento, Carla Carvalho, Don Colladay, Luis Guisado, Alan Kosteleck\'y, 
Clovis de Matos, Pedro Martins, David Mota, 
Jorge P\'aramos, Robertus Potting, Nuno Santos, Anjan Sen, Pedro Silva and 
Martin Tajmar. 
I would like also to express my gratitude to Serge Reynaud who 
suggested me to participate in the Rencontres de Moriond, 
Gravitational Waves and Experimental Gravity, and to Jacques Dumarchez 
for the patience, 
support and for the organization of such a pleasant and profitable meeting.

\end{document}